\documentstyle[12pt,epsf,epic,curves]{article}
\title{
  {\vspace{-3cm} \normalsize \hfill
    \parbox{38mm}{MS-TPI-97-8 \\
                  cond-mat/9708212}  }\\[25mm]
  The Interface Tension of the Three-dimensional\\
  Ising Model in Two-loop Order
  }
\author{Peter Hoppe and Gernot M\"unster \\
        Institut f\"ur Theoretische Physik I,
        Universit\"at M\"unster\\
        Wilhelm-Klemm-Str.~9, D-48149 M\"unster, Germany\\
        e-mail: munsteg@uni-muenster.de}
\date{August 27, 1997}
\newcommand{\be}{\begin{equation}}
\newcommand{\ee}{\end{equation}}
\newcommand{\bea}{\begin{eqnarray}}
\newcommand{\eea}{\end{eqnarray}}
\newcommand{\prop}{
\setlength{\unitlength}{1.5mm}
\begin{picture}(10,5)
\drawline(0,0.5)(10,0.5)
\end{picture}
}

\newcommand{\frprop}{
\setlength{\unitlength}{1.5mm}
\begin{picture}(7,5)
\dashline{1.0}(-3,0.5)(7,0.5)
\end{picture}
}

\newcommand{\drvert}{
\setlength{\unitlength}{1.5mm}
\begin{picture}(11,6)
\drawline(0,0.5)(6,0.5)
\drawline(6,0.5)(10,3.5)
\drawline(6,0.5)(10,-2.5)
\drawline(6,0.5)(6,-3.0)
\drawline(5.5,-2.5)(6.5,-3.5)
\drawline(5.5,-3.5)(6.5,-2.5)
\end{picture}
}

\newcommand{\drfrvert}{
\setlength{\unitlength}{1.5mm}
\begin{picture}(7,10)
\dashline[50]{1.0}(-2,0.5)(4.0,0.5)
\dashline[40]{1.0}(4.0,0.5)(8.0,3.5)
\dashline[40]{1.0}(4.0,0.5)(8.0,-2.5)
\end{picture}
}

\newcommand{\viervert}{
\setlength{\unitlength}{1.5mm}
\begin{picture}(4,8)
\drawline(-5.5,4.5)(2.5,-3.5)
\drawline(-5.5,-3.5)(2.5,4.5)
\end{picture}
}

\newcommand{\vierfrvert}{
\setlength{\unitlength}{1.5mm}
\begin{picture}(7,10)
\dashline[40]{1.0}(-1.5,4.5)(6.5,-3.5)
\dashline[40]{1.0}(-1.5,-3.5)(6.5,4.5)
\end{picture}
}

\newcommand{\nullmodvert}{
\setlength{\unitlength}{1.5mm}
\begin{picture}(8,7)
\put(1,0.5){\circle{2}}
\put(1,0.5){\circle*{1}}
\drawline(2,0.5)(7,0.5)
\end{picture}
}

\newcommand{\grapha}{
\setlength{\unitlength}{1.2mm}
\begin{picture}(10,4)
\put(2.5,0.5){\bigcircle{5}}
\put(7.5,0.5){\bigcircle{5}}
\end{picture}
}

\newcommand{\frgrapha}{
\setlength{\unitlength}{1.2mm}
\begin{picture}(10,4)
\curvedashes[0.8mm]{0,1,1.7}
\put(2.5,0.5){\bigcircle{5}}
\put(7.5,0.5){\bigcircle{5}}
\end{picture}
}

\newcommand{\graphb}{
\setlength{\unitlength}{1.3mm}
\begin{picture}(12,5)
\put(2,0.5){\bigcircle{4}}
\drawline(4,0.5)(8,0.5)
\drawline(4,0.5)(4,-3.5)
\drawline(3.5,-3.0)(4.5,-4.0)
\drawline(3.5,-4.0)(4.5,-3.0)
\put(10,0.5){\bigcircle{4}}
\drawline(8,0.5)(8,-3.5)
\drawline(7.5,-3.0)(8.5,-4.0)
\drawline(7.5,-4.0)(8.5,-3.0)
\end{picture}
}

\newcommand{\frgraphb}{
\setlength{\unitlength}{1.3mm}
\begin{picture}(12,3)
\curvedashes[0.8mm]{0,1,1.7}
\put(2,0.5){\bigcircle{4}}
\dashline[50]{0.8}(4,0.5)(8,0.5)
\put(10,0.5){\bigcircle{4}}
\end{picture}
}

\newcommand{\graphc}{
\setlength{\unitlength}{1.2mm}
\begin{picture}(10,5)
\curve(0,0.5,1,2.6,2,3.3,3,3.7,4,3.9,5,4,6,3.9,7,3.7,8,3.3,9,2.6,10,0.5)
\drawline(0,0.5)(10,0.5)
\curve(0,0.5,1,-1.6,2,-2.3,3,-2.7,4,-2.9,5,-3.0,6,-2.9,7,-2.7,8,-2.3,9,
        -1.6,10,0.5)
\drawline(0,0.5)(0,-4.5)
\drawline(-0.5,-4.0)(0.5,-5.0)
\drawline(-0.5,-5.0)(0.5,-4.0)
\drawline(10,0.5)(10,-4.5)
\drawline(9.5,-4.0)(10.5,-5.0)
\drawline(9.5,-5.0)(10.5,-4.0)
\end{picture}
}

\newcommand{\frgraphc}{
\setlength{\unitlength}{1.2mm}
\begin{picture}(10,5)
\curvedashes[0.8mm]{0,1,1.7}
\curve(0,0.5,1,2.9,2,3.7,3,4.2,4,4.4,5,4.5,6,4.4,7,4.2,8,3.7,9,2.9,10,0.5)
\dashline[50]{0.8}(0,0.5)(10,0.5)
\curve(0,0.5,1,-1.9,2,-2.7,3,-3.2,4,-3.4,5,-3.5,6,-3.4,7,-3.2,8,-2.7,9,-1.9,10,0.5)
\dashline[50]{0.8}(0,0.5)(10,0.5)
\end{picture}
}

\newcommand{\klgraphcmod}{
\setlength{\unitlength}{0.75mm}
\begin{picture}(10,5)
\curve(0,1.9,1,4.0,2,4.7,3,5.1,4,5.3,5,5.4,6,5.3,7,5.1,8,4.7,9,4.0,10,1.9)
\drawline(0,1.9)(10,1.9)
\curve(0,1.9,1,-0.2,2,-0.9,3,-1.3,4,-1.5,5,-1.6,6,-1.5,7,-1.3,8,-0.9,9,
        -0.2,10,1.9)
\drawline(0,1.9)(0,-3.1)
\drawline(-0.5,-2.6)(0.5,-3.6)
\drawline(-0.5,-3.6)(0.5,-2.6)
\drawline(10,1.9)(10,-3.1)
\drawline(9.5,-2.6)(10.5,-3.6)
\drawline(9.5,-3.6)(10.5,-2.6)
\end{picture}
}

\newcommand{\graphd}{
\setlength{\unitlength}{1.4mm}
\begin{picture}(12,7)
\put(1,0.5){\bigcircle{2}}
\put(1,0.5){\circle*{1}}
\drawline(2,0.5)(6,0.5)
\put(8,0.5){\bigcircle{4}}
\drawline(6,0.5)(6,-3.5)
\drawline(5.5,-3.0)(6.5,-4.0)
\drawline(5.5,-4.0)(6.5,-3.0)
\end{picture}
}
\begin{document}
\maketitle

\begin{abstract}
In liquid mixtures and other binary systems at low temperatures
the pure phases may coexist, separated by an interface.
The interface tension vanishes according to
$\sigma = \sigma_0 (1 - T/T_c)^{\mu}$ as the temperature $T$ approaches
the critical point from below.
Similarly the correlation length diverges as
$\xi = f_- (1 - T/T_c)^{-\nu}$ in the low temperature region.
For three-dimensional systems the dimensionless product
$R_- = \sigma_0 f_-^2$ is universal.
We calculate its value in the framework of field theory in $d=3$
dimensions by means of a saddle-point expansion around the kink solution
including two-loop corrections.
The result $R_- = 0.1065(9)$, where the error is mainly due to the
uncertainty in the renormalized coupling constant, is compatible with
experimental data and Monte Carlo calculations.\\[5mm]
\end{abstract}
%
For a statistical system near a critical point various measurable
quantities $X$ obey a singular behaviour as a function of the
temperature $T$ like
\be
X \sim X_0 t^{\varepsilon}
\ee
with a critical exponent $\varepsilon$, where
\be
t = \left|\frac{T-T_c}{T_c}\right| ,
\ee
and $T_c$ is the critical temperature.
For a given observable $X$ the critical exponent is a universal
quantity and assumes the same value for systems belonging to the same
universality class.
The critical amplitude $X_0$, however, is not universal and varies from
system to system, depending on the microscopic details of the
Hamiltonian.

The critical exponents of a universality class are not independent of each
other but obey a number of scaling and hyperscaling relations. For example,
in the low temperature region the exponents $\beta$, $\gamma$ and $\nu$
belonging to the magnetization
\be
M \sim B t^{\beta} ,
\ee
the susceptibility
\be
\chi \sim C_- t^{-\gamma} ,
\ee
and the correlation length
\be
\xi \sim f_- t^{-\nu}
\ee
are related by
\be
2\beta + \gamma - d\nu = 0 ,
\ee
where $d=3$ is the number of dimensions.
Therefore in the combination
\be
\label{uR}
u_R = \frac{3 \chi}{M^2 \xi^3}
\ee
the exponents cancel and $u_R$ is expected to approach a finite value
\be
u_R (t) \mathrel{\mathop{\longrightarrow}_{\scriptstyle {t \to 0}}} u_R^*
\ee
when the critical point is approached from below.
It is a another consequence of the scaling hypothesis that such
combinations of critical amplitudes are also universal
\cite{Fisk,Stauffer}.
They are generally called amplitude ratios.
For a review see \cite{PHA91}.

In the last decades much interest has focussed on critical indices,
whose values are known from various methods rather accurately by now.
From a phenomenological point of view the amplitude ratios are, however,
at least as interesting as the indices.
They are well accessible experimentally and their numerical values are
often more characteristic for the universality classes as the variation
between different classes are larger.

In this article we consider an amplitude ratio related to the interface
tension in the universality class of the three-dimensional Ising model.
In various binary systems at temperatures $T$ below $T_c$ interfaces
(domain walls) may be present, separating coexisting phases.
The interface tension $\tau$ is the free energy per unit area of
interfaces.
As $T$ increases towards $T_c$ the reduced interface tension
\be
\sigma = \frac{\tau}{kT} ,
\ee
where $k$ is Boltzmann's constant, vanishes according to the scaling law
\be
\sigma \sim \sigma_0 t^{\mu} .
\ee
Widom's scaling law \cite{Widom2,Widom1},
\be
\mu = 2\nu ,
\ee
relates the universal critical exponent $\mu$ to the critical exponent
$\nu$ of the correlation length $\xi$.
Associated with this law is the universal dimensionless product of
critical amplitudes
\be
R_- = \sigma_0 f_-^2 .
\ee

In this article we consider the correlation length as defined by means
of the second moment of the correlation function \cite{Tarko}, in
contrast to the ``true'' (exponential) correlation length.
Numerically they differ by less than 2 percent \cite{Caselle}.

The amplitude $\sigma_0$ has been studied experimentally (see
\cite{PHA91,Mainzer}) as well as theoretically.
The presently most accurate Monte Carlo result has been obtained by
Hasenbusch and Pinn \cite{HP}.

In the theoretical toolbox we also have the three-dimensional Euclidean
$\phi^4$-theory, which is believed to be in the same universality class
as binary systems and the Ising model.
Therefore it should describe the universal properties of these systems
correctly.
The scalar field $\phi(x)$ represents the local order parameter which
in the case of binary fluid mixtures is proportional to the difference
of the concentrations of the two fluids.

On the classical level an interface in a system of cylindrical geometry
is represented by a classical solution of the field equations.
It is a saddle-point of the Hamiltonian.
Taking thermal fluctuations into account amounts to performing an
expansion around the saddle-point in the functional integral.

In the field theoretical framework the interface tension, in
particular the universal ratio $R_-$, has been investigated
by means of the $\epsilon$-expansion in $4-\epsilon$ dimensions
\cite{Pant,BF} and directly in 3 dimensions \cite{Mue3d}.
In both approaches the calculations were done in the one-loop
approximation (see below).
Whereas the results from the $\epsilon$-expansion are afflicted by
convergence problems and show large deviations from experimental and
Monte Carlo results, the three-dimensional field theory leads to
relatively small one-loop corrections and more reasonable numbers.
But higher-loop corrections may spoil this situation.
Therefore, in order to get a better impression of the numerical
convergence and to obtain more precise estimates it is highly desirable
to know the two-loop contribution to $R_-$.
In this article we present the result of a two-loop calculation of the
universal amplitude ratio $R_-$ in the framework of three-dimensional
$\phi^4$-theory.

The Hamiltonian $\cal H$, which is called action in the context of
Euclidean field theory, in the broken symmetry phase is written in terms
of the bare field $\phi_0$ as
\be
{\cal H} = \int\!\!{\cal L}\,d^3x , \hspace{1cm}
{\cal L} = \frac{1}{2} \partial_{\mu} \phi_0 \,\partial^{\mu} \phi_0
           + V(\phi_0) ,
\ee
where the double-well potential
\be
V(\phi_0) = - \frac{m_0^2}{4} \phi_0^2 + \frac{g_0}{4!} \phi_0^4
            + \frac{3}{8} \frac{m_0^4}{g_0}
= \frac{g_0}{4!} \left( \phi_0^2 - v_0^2 \right)^2
\ee
has its minima at
\be
\phi_0 = \pm v_0 = \pm \sqrt{\frac{3 m_0^2}{g_0}} .
\ee
The parameters are defined such that the value of the potential at
its minima is zero and $m_0$ is the bare mass.
The renormalized mass
\be
\label{mR}
m_R = 1/\xi
\ee
is the inverse of the second moment correlation length.
It is defined together with the wave function
renormalization $Z_R$ through the small momentum behaviour of the
propagator:
\be
G(p)^{-1} = \frac{1}{Z_R} \{ m_R^2 + p^2 + O(p^4) \}.
\ee
The renormalized vacuum expectation value of the field is
\be
v_R =  Z_R^{-1/2} \, v ,
\ee
where $v$ is the expectation value of the field $\phi_0$.
For the dimensionless renormalized coupling we adopt the definition
\be
u_R = \frac{g_R}{m_R} = 3\,\frac{m_R}{v_R^2} ,
\ee
which in the language of statistical mechanics corresponds to
Eq.~(\ref{uR}).

The basic idea behind the calculation of the interface tension is its
relation to the energy splitting due to tunneling in a finite volume.
We refer to \cite{Mue4d,Mue3d} for details.
In a rectangular box with cross-section $L^2$ the degeneracy of the
groundstate of the transfer matrix is lifted by an ``energy'' splitting
$E_{0a}$.
This gap depends on $L$ according to
\be
E_{0a} = C \exp \left\{ -\sigma L^2 \right\} ,
\ee
where $\sigma$ is the (reduced) interface tension
\cite{FISHER,PF,BZ,Mue3d}.

The energy splitting $E_{0a}$ can be calculated in a semiclassical
approximation, which amounts to a saddle-point expansion around the
classical kink solution
\be
\phi_c(x) = \sqrt{\frac{3m_0^2}{g_0}} \tanh \frac{m_0}{2} (x_3 - a)
\ee
of $\phi^4$-theory, where $a$ is a free parameter specifying the
location of the kink.
The classical energy of a kink is
\be
{\cal H}_c = 2 \,\frac{m_0^3}{g_0} L^2 .
\ee
The kink interpolates between the two field values at the minima of the
potential and represents an interface separating regions with different
local mean values of the field.

In the two-loop approximation the functional integral
\be
Z_{+-} = \int\!e^{-{\cal H}[\phi_0]}\,{\cal D}\phi_0
\ee
(a factor of $1/kT$ has been absorbed in $\cal H$)
with appropriate boundary conditions is calculated by expanding the
energy ${\cal H}[\phi]$ around the kink solution $\phi_c$ up to order
$g_0$ and evaluating the integral by the saddle point method.
For details see \cite{HM}.
An analogous calculation for the case of the anharmonic oscillator in
quantum mechanics has been performed in \cite{Shuryak}.

The zero mode belonging to the translations of the center of the kink,
i.e.\ to shifts in $a$ is treated by the method of collective
coordinates and leads to a nontrivial Jacobian $J$.
For the energy splitting one obtains
\begin{eqnarray}
E_{0a}&=&2\,\sqrt{\frac{{\cal H}_c}{2 \pi}}\;
            \bigg(\frac{\det'M}{\det M_0}\bigg)^{-1/2}
\nonumber\\
&&\times\,
       \exp\Bigg\{-{\cal H}_c\;+\;\frac{1}{2}\;\graphd\;+\;
        \frac{1}{8}\;\bigg[\;\grapha\;-\;\frgrapha\;\bigg]
\nonumber\\
&&\hspace{1.45cm}+\;\frac{1}{8}\;\bigg[\;\graphb\;-\;\frgraphb\;\bigg]
    \;+\;\frac{1}{12}\;\bigg[\;\graphc\;-\;\frgraphc\;\bigg]
\nonumber\\[1.5ex]
&&\hspace{1.45cm}
      \;+\; O\big(g_0^2\big)\Bigg\} ,
\end{eqnarray}
where
\be
M = - \partial_{\mu} \partial^{\mu} + m_0^2
   - \frac{3}{2} m_0^2 \cosh^{-2}\left(\frac{m_0}{2} x_3\right) 
\ee
is the operator of quadratic fluctuations around $\phi_c$ and
\be
M_0 = - \partial_{\mu} \partial^{\mu} + m_0^2 .
\ee
The prime in $\det'$ indicates a determinant without zero-modes.
The two-loop contributions are displayed as Feynman diagrams.
The propagators are
\begin{eqnarray}
\prop&\widehat{=}&G(x,y) ,
\nonumber\\
\frprop&\widehat{=}&G_0(x,y) ,
\end{eqnarray}
where the propagator in the kink background
\be
G(x,y) = \langle x| (M')^{-1} |y \rangle
\ee
is the Greens function of $M$ without zero mode and
\be
G_0 (x,y) = \langle x| M_0^{-1} |y \rangle
\ee
is the usual scalar propagator.
Remember, however, that the propagators refer to a system with finite
cross-section $L^2$.

The vertices are
\begin{eqnarray}
\drvert&\widehat{=}& - g_0\,\phi_c(x)\;
            =\;-\;\sqrt{3\,g_0\,m_0^2}\;
                  \tanh \frac{m_0}{2} x_3 
\nonumber\\
\viervert&\widehat{=}& -\;g_0 ,
\nonumber\\
\drfrvert&\widehat{=}&  -\;\sqrt{3\,g_0\,m_0^2} ,
\nonumber\\
\vierfrvert&\widehat{=}& -\;g_0 ,
\nonumber\\
\nullmodvert&\widehat{=}&-\frac{1}{{\cal H}_c}\;\ddot{\phi}_c(x) .
\end{eqnarray}
The last vertex comes from the Jacobian $J$.

The spectrum of the fluctuation operator $M$ is known exactly.
Owing to this the determinants can be evaluated analytically with the
help of heat kernel and zeta-function techniques \cite{Mue4d,Mue3d}.
They yield the one-loop contribution to the interface tension.

Much more involved is the calculation of the two-loop contributions.
Although we have an expression for $G(x,y)$ (covering one and a half
page) it turned out to be advantageous to use the Schwinger
representation
\be
G = \int_0^{\infty}\!e^{-t M'}\,dt
\ee
and to write the kernel $\exp(-tM')$ in the spectral representation.
The calculations have been done analytically as far as possible.
In the later stages some infinite sums and low-dimensional integrations
have been done numerically. Details of the calculation can be found in
\cite{HM}.

The most difficult piece, of course, was the true two-loop diagram
$\;\rule[-0.25cm]{0.0cm}{0.7cm}\klgraphcmod\;$.
It contains ultraviolet divergencies, which we isolated by means of
dimensional regularization in $d=3-\epsilon$ dimensions.
After some tedious calculations (we warn the curious reader) we obtained
the two-loop contribution as a function of $L$ up to terms of order
$L^0$.
Whereas individual diagrams produce terms of the form
$L^2 (\log(m_0 L))^2$ and $L^2 \log(m_0 L)$, they cancel in the total
sum.
The leading $L$-dependence is then proportional to $L^2$ as is required
by the finiteness of the interface tension.

The ultraviolet divergence is removed by renormalization as usual.
Expressing the bare parameters $m_0$ and $g_0$ in terms of their
renormalized counterparts $m_R$ and $g_R$ in the two-loop approximation
indeed cancels the pole in $1/\epsilon$.
The final result for the interface tension at $L = \infty$ is
\be
\sigma = \frac{2 m_R^2}{u_R} \left\{ 1 + \sigma_{1l}\,\frac{u_R}{4\pi}
+ \sigma_{2l}\left(\frac{u_R}{4\pi}\right)^2 + O(u_R^3) \right\} ,
\ee
with
\be
\sigma_{1l} =
\frac{1}{4} \left( 3 + \frac{3}{4} \log 3 \right) - \frac{37}{32}
= -0.2002602
\ee
and
\be
\sigma_{2l} = -0.0076(8) .
\ee

Using (\ref{mR}) the desired amplitude ratio $R_-$ is obtained by
evaluating the function
\be
f(u_R) = \sigma / m_R^2
\ee
at the fixed point value $u_R = u_R^*$, i.e.
\be
R_- = f(u_R^*) =
\frac{2}{u_R^*} \left\{ 1 + \sigma_{1l}\,\frac{u_R^*}{4\pi}
+ \sigma_{2l}\left(\frac{u_R^*}{4\pi}\right)^2 + O(u_R^{* 3}) \right\} .
\ee
The most recent results for $u_R^*$ are
\be
u_R^* = 14.3(1)
\ee
from Monte Carlo calculations \cite{Caselle} and $u_R^* \approx 14.2$
from three-dimensional field theory \cite{GKM}.
Earlier estimates were 14.73(14) from low-temperature series
\cite{Siepmann} and 15.1(1.3) used in \cite{Mue3d}.
For these numbers the two-loop contribution to $R_-$ is about 1\,\%
while the one-loop contribution is about 24\,\%.
Although the apparent numerical convergence is surprisingly good we have
also applied Pad\'{e} and Pad\'{e}-Borel approximations to the quadratic
polynomial appearing in $f(u_R)$ in order to get $R_-$.
The dependence of $R_-$ on $u_R^*$ is illustrated in Fig.~1, where the
average of the three Pad\'{e} approximants and the corresponding error
band is displayed.
\begin{figure}[t]
\centering
\epsfxsize 15cm
\epsffile{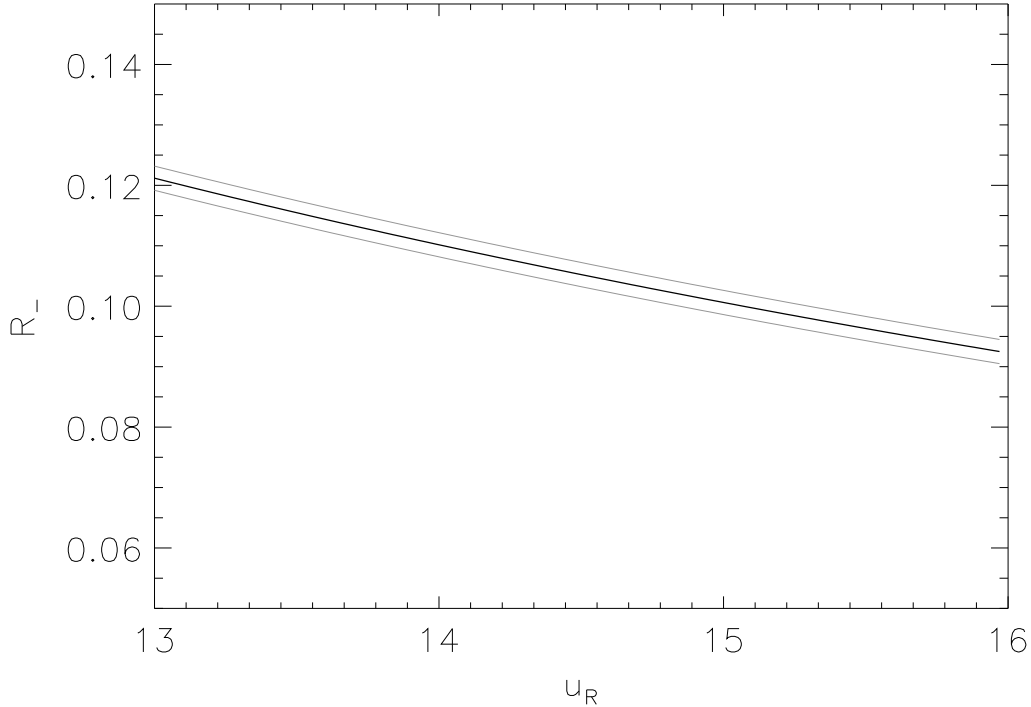}
\parbox[t]{0.8\textwidth}{
\caption{$R_-$ as a function $u_R^*$. The solid line is the average of the
three Pad\'{e} approximants. The thin lines indicate an error estimate of
0.001.}
}
\end{figure}

Table 1 shows the results of the Pad\'{e} and Pad\'{e}-Borel
approximations evaluated at three of the values for $u_R^*$ mentioned
above.
The quoted errors are due to the error of $\sigma_{2l}$.
\begin{table}[ht]
\begin{center}
\begin{tabular}{|c|c|c|c|c|c|}
\hline
&\multicolumn{5}{|c|}{$R_-$}\\[0.06cm]
\hline
$u_R^*$&$f$&$f_{[1,1]}$&$f_{[0,2]}$&
   $f_{[1,1],\mbox{\tiny PB}}$&$f_{[0,2],\mbox{\tiny PB}}$\\[0.06cm]
\hline
15.1&0.0991(2)&0.0989(2)&0.1011(1)&0.0991(3)&0.10454(5)\\
\hline
14.73&0.1025(2)&0.1024(2)&0.1044(1)&0.1025(3)&0.10774(5)\\
\hline
14.3&0.1066(2)&0.1064(2)&0.1084(1)&0.1066(3)&0.11166(5)\\
\hline
\end{tabular}
\parbox[t]{0.8\textwidth}{
\caption{Results for $R_-$ from Pad\'{e} and Pad\'{e}-Borel              
approximations evaluated at three values for $u_R^*$.}
}
\end{center}
\end{table}

The average of all approximants at $u_R^* = 14.3$ is $R_- = 0.108(2)$.
The [0,2] approximants appear to be off the rest.
Leaving them out yields $R_- = 0.1065(1)$.
Taking the error of $u_R^*$ into account we obtain
\be
R_- = 0.1065(9) .
\ee
Since we do not know the size of higher-loop contributions
the quoted error mainly reflects the spread due to the uncertainty of
$u_R^*$.

For comparison the Monte Carlo calculations of Hasenbusch and Pinn
\cite{HP} yield
\be
R_- = 0.1040(8) .
\ee
In view of the remarks made above we consider the results as being
compatible.

Experimentally the universal amplitude combination $R_+ = \sigma_0 f_+^2$,
where $f_+$ is the amplitude of the correlation length in the high
temperature phase, has been measured for various binary systems, see
\cite{PHA91}.
In order to compare with $R_-$ the universal ratio $f_+ / f_-$ has to be
employed.
Using $f_+ / f_- = 1.95(2)$ from recent Monte Carlo calculations
\cite{Caselle} or $f_+ / f_- = 1.99(2)$ from field theory (see \cite{GKM}
and the remark in the conclusions of \cite{Caselle}) we obtain for $R_+$ the
numbers 0.40(1) and 0.42(1), respectively.
This compares well with the recent experimental result of 0.41(4) for the
classical cyclohexane-aniline mixture \cite{Mainzer}.
Previous experimental results are summarized in \cite{Mainzer} as 0.37(3).


%
\end{document}